\documentclass[aps,prd,twocolumn,showpacs,preprintnumbers,amsmath,amssymb]{revtex4-1}

\usepackage{graphicx}
\usepackage[usenames,dvipsnames]{xcolor}
\usepackage{braket}
\usepackage{amssymb}
\usepackage{amsfonts}
\usepackage{amsmath}
\usepackage{bbm} 
\usepackage{bbold}
\usepackage{siunitx}

\usepackage{xargs}      
\usepackage[colorlinks]{hyperref}
\usepackage{makeidx}
\definecolor{winered}{rgb}{0.8,0,0}
\definecolor{darkb}{rgb}{0,0,0.8}
 \usepackage{hyperref}
\hypersetup{
    colorlinks=true,
    citecolor=blue,
    linkcolor=winered,
    filecolor=red,      
    urlcolor=darkb,
}
 
\usepackage{lineno}
  \usepackage{young}
\begin{document}
\title{Mass splittings in a linear sigma model for multiflavor gauge theories}
\author{D. Floor$^1$}
\author{E. Gustafson$^1$}
\author{Y. Meurice$^1$}
\affiliation{$^1$ Department of Physics and Astronomy, The University of Iowa, Iowa City, IA 52242, USA }
\definecolor{burnt}{cmyk}{0.2,0.8,1,0}
\def\lt{\lambda ^t}
\def\note{note}
\def\beq{\begin{equation}}
\def\enq{\end{equation}}
\newcommand{\Tr}{\text{Tr}}

\def\gf{U(N_f)_L\bigotimes U(N_f)_R}
\def\ls{\lambda_\sigma}
\def\la{\lambda_{a0}}
\def\pdp{\phi^\dagger\phi}
\def\bpi{\boldsymbol\pi}
\def\bao{\bf a_0}
\def\etp{\eta'}
\date{\today}
\begin{abstract}
We calculate the tree-level mass spectrum for a linear 
sigma model describing the scalar and pseudoscalar mesons of 
a $SU(3)$ local gauge theory with Dirac fermions in the fundamental representation. $N_1$ fermions have a  mass $m_1$ and $N_2$  a mass $m_2$. Using recent lattice data with $m_1=m_2$ and $N_1+N_2$= 8  or 12, 
we predict the mass splittings for $m_2=m_1+\delta m$. At first order in $\delta m$, an interesting inverted pattern appears in 
the $0^{++}$ sector, where mesons with lighter fermions are heavier. This feature could be tested in ongoing calculations provided that $m_1$ and $\delta m$ are sufficiently small. We discuss possible improvements of the approach.
\end{abstract}

\maketitle
\section{Introduction}
The idea that asymptotically free gauge theories with a sufficiently large number of massless fermions (flavors) have a nontrivial infrared fixed point (IRFP) 
has motivated many lattice studies \cite{bsr,dgrmp,nogrev}. 
It is expected that in the massless limit, conformal symmetry and 
gapless deconfined excitations are present.
This limiting situation is unlikely to be directly relevant for particle physics. Nevertheless by introducing mass terms or reducing 
the number of flavors slightly below a critical value, one can obtain possibly interesting models  in the context of electroweak symmetry breaking. 
If an IRFP exists, fermion masses provide relevant directions out of this IRFP which are expected to drive the renormalization group (RG) flows towards fixed points where a 
more conventional behavior is expected. However, a consensus on a physical picture supported by an effective theory is still lacking \cite{bsr,dgrmp,nogrev}. 

In this article we focus on the well-studied example of a $SU(3)$ gauge theory with $N_f$ fundamental Dirac fermions. In the massless limit, we have a clear physical picture when the $N_f$ is not too large, say for $N_f \leq 4$, the clearly QCD-like region: there are $N_f^2-1$ massless pions and the other states  (scalars with positive parity, baryons, ...) are massive. At the other end, for $N_f=16$, the last value preserving asymptotic freedom, the two-loop beta function has a non-trivial zero at $\alpha_c\simeq 1/20$ and perturbation theory should be valid to describe weakly interacting massless deconfined quarks and gluons. 
It is clear that as $N_f$ is increased between these two limits, the low energy degrees of freedom change drastically, however a consensus on the details of the 
changes is not available so far. One important limitation is that lattice simulations with low fermion masses are typically impractical and one has to rely on models for the massless extrapolation. Nevertheless there is a consensus on the fact that adding light flavors tend 
to produce unexpectedly light states besides the pions. 

Light $\sigma$ masses were found for $SU(3)$ gauge theories with 
8 \cite{Aoki:2014oha,Appelquist:2016viq,Aoki:2016wnc,Gasbarro:2017fmi} and 12 \cite{Aoki:2013zsa,Aoki:2016wnc} fundamental flavors and also for 2 sextets \cite{Fodor:2015vwa}. 
Recent results \cite{latkmi16,Aoki:2017fnr} concerning the mass of the $\etp$ support the possibility \cite{meuriceprd96} that the explicit breaking of the axial $U(1)_A$ symmetry, which depends in a distinct way on $N_f$, can explain the fact that the $\sigma$ become lighter as $N_f$ increases. One simple and interesting possibility \cite{bz,st05,ds07} is that at some point $N_f$ reaches the boundary of the conformal window which is signaled by the fact that the $\sigma$ and other states become massless. However, more complex intermediate situations or phases are conceivable when $N_f$ is increased. 

An interesting question is to figure out if unexpectedly light states persist when a certain number of fermions have a larger mass. 
In the following, we consider a linear sigma model with $N_1$ light hyperquarks of mass $m_1$ and $N_2$ heavier hyperquarks of mass $m_2$. This model is an extension of the single mass model discussed in Ref. \cite{meuriceprd96} and which was introduced and studied in QCD context  \cite{PhysRevD.3.2874,PhysRevD.21.3388,tHooftphysrep,Meuricea0,PhysRevD.77.094004,PhysRevD.82.054024,PhysRevD.87.014011,Kovacs:2013xca,PhysRevD.75.025015,PhysRevD.93.114014}. The case $N_1=2$, $N_2\geq 4$ 
could provide interesting extensions of the minimal technicolor scenario \cite{csakirmp} but it has not been studied on the lattice so far because multiples of 4 are convenient with staggered fermions. 
Part of the spectrum for $N_1=4$, $N_2=8$ has been extracted from recent lattice simulations \cite{Hasenfratz:2016gut,PhysRevD.93.075028}.  A possible phenomenological motivation for this choice is given in Ref. \cite{Ma:2015gra}.

The article is organized as follows. The linear sigma model is presented in Sec. \ref{sec:model}. The tree-level spectrum is calculated in Sec. \ref{sec:spectrum}. In the one fermion mass case, 
$2N_f^2$  bosons are characterized by 4 masses ($\sigma$, $\bao$, $\etp$ and $\bpi$). With two fermion masses, the $\bao$ and $\bpi$
each split into four representations of the unbroken subgroup $SU(N_1)_V\bigotimes  SU(N_2)_V$. Sec. \ref{sec:unpert} discusses 
the determination of the parameters of the model in terms of the masses in the equal mass case.  We emphasize that if we want to fit the spectrum using the tree level model, 
the quartic couplings depend on the symmetry breaking term in a way that is only understood empirically. In Sec. \ref{sec:pert}
we introduce a perturbative approach where $m_2=m_1+\delta m$ and calculate the mass splitting at first order in $\delta m$. 
This allows us to use the empirical unperturbed (single mass) results \cite{Aoki:2016wnc,Aoki:2017fnr} to estimate the mass splittings. 
We obtain simple ratios of differences between masses squared which are identical for scalar and pseudoscalar. However, the numerical results of the LatKMI collaboration \cite{Aoki:2016wnc,Aoki:2017fnr} indicate that at first order there is an interesting inversion for the adjoint $0^{++}$ (the $\bao$), namely the 
meson containing light hyperquarks are heavyier than those containing one or two heavy hyperquarks. In the conclusions, we discuss how to test these predictions in ongoing lattice simulations.

\section{The two mass model}
\label{sec:model}
The model considered here is introduced and motivated in Refs. \cite{PhysRevD.3.2874,PhysRevD.21.3388,tHooftphysrep,Meuricea0,meuriceprd96}. The only difference with Ref. \cite{meuriceprd96} is the explicit breaking of the chiral symmetry which here corresponds to $N_1$ hyperquarks of mass $m_1$ and $N_2$ hyperquarks of mass $m_2$, with 
$N_1+N_2=N_f$. 
For the sake of self-containedness, some basic points are repeated below. 

The effective fields $\phi_{ij}$ are $N_f\times N_f$ matrices transforming as  $\bar{\psi}_{Rj}\psi_{Li}$ under $\gf$. We use the parametrization.
\beq
\label{eq:phip}
\phi=(S_\alpha+iP_\alpha)\Gamma^\alpha, 
\enq
with a summation over $\alpha=0,1,\dots N_f^2-1$ for a basis of $N_f\times N_f$ Hermitian matrices $\Gamma^\alpha$ such that 
\beq
\label{eq:norm}
Tr(\Gamma^\alpha\Gamma^\beta)=(1/2)\delta^{\alpha\beta}, 
\enq
We use the convention that $\Gamma_0=\mathbb{1}_{N_f\times N_f}/\sqrt{2N_f}$ while the remaining $N_f^2-1$ matrices are traceless. 
The $S_0$ and $P_0$ correspond to the $\sigma$ and $\eta'$ respectively  while the remaining components transform like the adjoint representation and are denoted ${\bao}$ and $\bpi$ respectively. In addition, we define a $N_f\times N_f$ matrix for which we use the short notation $``\Gamma^8 "$ in analogy with the 2+1 flavors case and which is defined as 
\begin{equation}
 \Gamma^8\equiv\frac{1}{\sqrt{2 N_f}}
\begin{pmatrix}
    \sqrt{N_2/N_1}\  \mathbb{1}_{N_1 \times N_1}& 0\\
0 & -\sqrt{N_1/N_2}\  \mathbb{1}_{N_2 \times N_2}\\
\end{pmatrix} 
\end{equation}

The effective Lagrangian has a canonical kinetic term 
\beq
\label{eq:lag}
{\mathcal{L}}_{kin.}=Tr\partial_\mu\phi\partial^\mu\phi^\dagger, 
\enq
and a potential term consisting of  three parts 
\beq
V=V_0+V_a+V_m. 
\enq
The first two terms are given as 
\begin{eqnarray}
\label{eq:vo}
V_0&\equiv&-\mu^2Tr(\phi^\dagger\phi)+(1/2)(\ls-\la)(Tr(\pdp))^2\nonumber\\
&\ &+(N_f/2)\la Tr((\pdp)^2). 
\end{eqnarray}
and 
\beq
V_a\equiv-2(2N_f)^{N_f/2-2}X(det\phi + det\phi^\dagger). 
\enq
The third term represent the effect of mass term with $N_1$ flavors  of mass $m_1$ and $N_2$ flavors of mass $m_2$,
\beq
V_m\equiv-(Tr\mathcal{M}\phi+h.c.)=-b_0 S_0-b_8S_8
\enq.
The matrix $\mathcal{M}$ can be written as $b_0 \Gamma^0 + b_8 \Gamma^8$ and $V_m$ is invariant under 
$SU(N_1)_V\bigotimes SU(N_2)_V$. We assume that this vector symmetry is not broken spontaneously and that the
vacuum expectation of $\phi$ has the form: 
\begin{equation}
\langle \phi \rangle = \frac{1}{\sqrt{2 N_f}}.
\begin{pmatrix}
    v_1 \mathbb{1}_{N_1 \times N_1}& 0\\
0 & v_2 \mathbb{1}_{N_2 \times N_2}.\\
\end{pmatrix}\end{equation}
This means that $\langle S_0\rangle =v_0$ and  $\langle S_8\rangle =v_8$ or equivalently 
\beq
\langle \phi \rangle = v_0 \Gamma^0 + v_8 \Gamma^8.
\enq
The transformation between the two expressions is 
\begin{eqnarray}
v_1&=&v_0+\sqrt{N_2/N_1} v_8\\
v_2&=&v_0-\sqrt{N_1/N_2} v_8,
\end{eqnarray}
and its inverse 
\begin{eqnarray}
v_0&=&(1/N_f)(N_1v_1+N_2v_2)\\
v_8&=&(\sqrt{N_1N_2}/N_f)(v_1-v_2).
\end{eqnarray}
The same transformation can be used to define $b_1$ and $b_2$ in terms of $b_0$ and $b_8$. This implies that 
\beq
b_0 S_0+b_8S_8=(N_1/N_f)v_1b_1+(N_2/N_f)v_2b_2.
\enq
The vacuum values $v_1$ and $v_2$ are given in terms of the couplings using the minimization conditions which read
\begin{eqnarray}
M_{\pi_{ll}}^2v_1 &=& b_1\\
M_{\pi_{hh}}^2 v_2&= &b_2, 
\end{eqnarray}
with $M_{\pi_{ll}}$ and $M_{\pi_{hh}}$, the pseudo-Nambu-Goldstone boson masses corresponding to light-light and 
heavy-heavy hyperquarks and discussed in the next section.
\section{The spectrum}
\label{sec:spectrum}
The spectrum of the model can be obtained from the second derivatives of the potential at $\langle \phi \rangle$:
\begin{equation}
\begin{split}
M^2_{S\alpha\beta}&\equiv \partial^2 V/\partial S_\alpha\partial S_\beta |_{\left\langle\phi \right\rangle}\\
M^2_{P\alpha\beta}&\equiv \partial^2 V/\partial P_\alpha\partial P_\beta |_{\left\langle\phi \right\rangle}.\\
\end{split}
\end{equation}
When the two masses are equal each parity sector splits into a singlet and the adjoint of $SU(N_f)_V$. 
We now consider the effect of having two masses with the convention $m_1\leq m_2$. 
We call the $N_1$ flavors  of mass $m_1$ ``light" and the $N_2$ flavors of mass $m_2$
``heavy". The adjoint of $SU(N_f)_V$ can be decomposed into representation of the $SU(N_1)_V\bigotimes SU(N_2)_V$
subgroup as follows:
\beq 
\label{eq:split}
(N_1^2-1,1)\bigoplus(1,N_2^2-1)\bigoplus ((N_1,\bar{N_2})+h.c.)\bigoplus(1,1).\nonumber\enq
We call the first three representations light-light ($l l$), heavy-heavy ($hh$), heavy-light ($hl$). The last one is the singlet associated with 
$\Gamma^8$. Except for one mixing between the indices 0 and 8,  $M^2_{S\alpha\beta}$ 
and $M^2_{P\alpha\beta}$ are diagonal. All the diagonal terms have a common term:
\beq
\mathcal{C} \equiv  - \mu^2 + \frac{\lambda_{\sigma} - \lambda_{a0}}{2 N_f}(N_1 v_1^2 + N_2 v_2^2).
\enq

We now proceed to give explicit expressions for the second derivatives. 
For the non-singlet pseudoscalars we have
\begin{equation}
\begin{split}
M_{\pi_{ll}}^2 &= \mathcal{C} +\frac{\lambda_{a0}}{2}v_1^2 - \frac{X}{N_f} v_1^{N_1 - 2} v_2^{N_2}\\
M_{\pi_{lh}}^2 -M_{\pi_{ll}}^2 &= \Big(\frac{\lambda_{a0}}{2} +  \frac{X}{N_f} v_1^{N_1-2}v_2^{N_2 - 2}\Big)(v_2-v_1)v_2\\
M_{\pi_{hh}}^2 - M_{\pi_{ll}}^2 &= \Big(\frac{\lambda_{a0}}{2} +  \frac{X}{N_f}v_1^{N_1 - 2}v_2^{N_2 - 2}\Big)(v_2^2 - v_1^2)\\.
\end{split}
\end{equation}
As explained above, the two singlets have a mixing term and the spectrum in this sector is given by the eigenvalues of a  $2\times2$ matrix.
The situation is similar to having the physical $\eta$ and $\eta'$ given as  mixings of the mathematical $SU(3)$ and $SU(2)\bigotimes U(1)$ singlets in three flavor QCD.   
\begin{equation}
\begin{split}
M_{P_{00}}^2 &= \mathcal{C}
 +\frac{\lambda_{a0}}{2 N_f}(N_1 v_1^2 + N_2 v_2^2) \\ + \frac{X}{N_f^2} &v_1^{N_1-2} v_2^{N_2-2}((N_1 v_2 + N_2 v_1)^2 - (N_1 v_2^2 + N_2 v_1^2))\\
M_{P_{88}}^2& =  \mathcal{C}
+\frac{\lambda_{a0}}{2 N_f}(N_2 v_1^2 + N_1 v_2^2) \\
+ \frac{X}{N_f^2} &v_1^{N_1-2} v_2^{N_2-2}(N_1 N_2 ( v_2 - v_1)^2 - (N_2 v_2^2 + N_1 v_1^2))\\
M_{P_{08}}^2 &=  (v_2-v_1)\Big[ -\frac{\lambda_{a0}}{2 N_f}\sqrt{N_1 N_2}(v_1+v_2)  \\
+ \frac{X}{N_f^2}& v_1^{N_1-2} v_2^{N_2-2}\sqrt{N_1 N_2}((N_1 v_2 + N_2 v_1) - (v_2 + v_1)) \Big]\\.
\end{split}
\end{equation}

For the non-singlet scalars, we have
\begin{equation}
\label{eq:scpsc}
\begin{split}
M^2_{a0_{ll}} - M_{\pi_{ll}}^2 &= \lambda_{a0} v_1^2 + \frac{2X}{N_f} v_1^{N_1 - 2} v_2^{N_2}\\
M^2_{a0_{lh}} - M_{\pi_{lh}}^2 &= \lambda_{a0} v_1 v_2+ \frac{2X}{N_f} v_1^{N_1 - 1} v_2^{N_2-1}\\
M^2_{a0_{hh}} - M_{\pi_{hh}}^2 &= \lambda_{a0} v_2^2 + \frac{2X}{N_f} v_1^{N_1} v_2^{N_2-2}\\.
\end{split}
\end{equation}
For the two singlets and their mixing 
\begin{equation}
\begin{split}
M_{S_{00}}^2 &= \mathcal{C} +\frac{\lambda_{\sigma} - \lambda_{a0}}{2 N_f}( \frac{2}{N_f}(N_1 v_1 + N_2 v_2)^2)\\& +\frac{3\lambda_{a0}}{2 N_f} (N_1 v_1^2 + N_2 v_2^2)\\ - \frac{X}{N_f^2} &v_1^{N_1-2} v_2^{N_2-2}((N_1 v_2 + N_2 v_1)^2 - (N_1 v_2^2 + N_2 v_1^2))\\
M_{S_{88}}^2 &= \mathcal{C} +\frac{\lambda_{\sigma} - \lambda_{a0}}{2 N_f}( \frac{2 N_1 N_2}{N_f}(v_1 - v_2)^2)\\ &+\frac{3\lambda_{a0}}{2 N_f} (N_2 v_1^2 + N_1 v_2^2)\\ - \frac{X}{N_f^2}& v_1^{N_1-2} v_2^{N_2-2}(N_1 N_2 ( v_2 - v_1)^2 - (N_2 v_2^2 + N_1 v_1^2))\\
M_{S_{08}}^2 &=(v_2-v_1)\big[-\frac{\lambda_{\sigma} - \lambda_{a0}}{N_f^2}(\sqrt{ N_1 N_2}(N_1 v_1 + N_2 v_2)) \\&-\frac{3\lambda_{a0}}{2 N_f} \sqrt{N_1 N_2}(v_1 + v_2)\\ - \frac{X}{N_f^2} &\sqrt{N_1N_2}v_1^{N_1-2} v_2^{N_2-2}((N_2 v_1 + N_1 v_2) - (v_2 + v_1))\big]\\.
\end{split}
\end{equation}
\section{Remarks about the unperturbed spectrum}
\label{sec:unpert}
Before applying the equations given in Sec. \ref{sec:spectrum} to practical situations, we need to clarify some aspects of 
our understanding of the unperturbed model, in other words, with $m_f=m_1=m_2$ and $N_f=N_1+N_2$ flavors. In the context of QCD it is possible to use chiral perturbation theory to calculate the 
way the masses of mesons and couplings change when small quark masses are modified by small amounts \cite{Scherer:2005ri,bijnensreview}. 
However, so far, this is not the case for $N_f=8$ or 12. 

It is commonly believed that for $N_f=8$ chiral symmetry is broken spontaneously, 
however the ratio $M_\pi^2/m_f$ has large nonlinear variations for $0.01<am_f<0.05$ compared to $N_f=4$ in the same $am_f$ range, $a$ being the lattice spacing. This can be seen clearly by comparing Figs. 22 and 23 in Ref. \cite{Aoki:2016wnc}. A detailed 
discussion of the applicability of chiral perturbation theory for $N_f=8$ can be found in Sec. IV of Ref. \cite{Aoki:2016wnc}, where 
it is stated that there is no numerical evidence for the predicted chiral logs. For $N_f=12$, the situation is more controversial. It is clear that if chiral symmetry is unbroken in the massless limit, the conventional tools are not useful. 

\begin{figure}[h]
\vskip-80pt
\includegraphics[width=9.6cm]{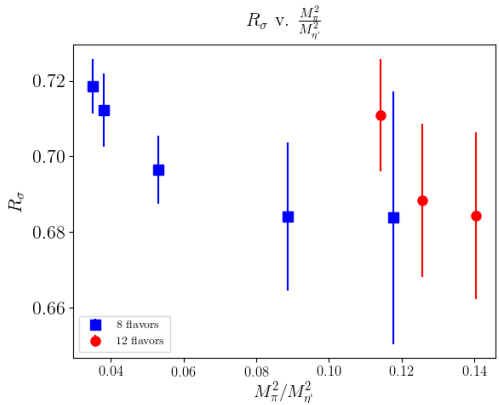}
\vskip-80pt
\caption{\label{fig:ratios}
$R_{\sigma}$  for $N_f$=8 (squares) and 12 (circles)   versus 
$(M_\pi/M_{\etp})^2$.}
\end{figure}

\begin{figure}[h]
\vskip-80pt
\includegraphics[width=7.6cm]{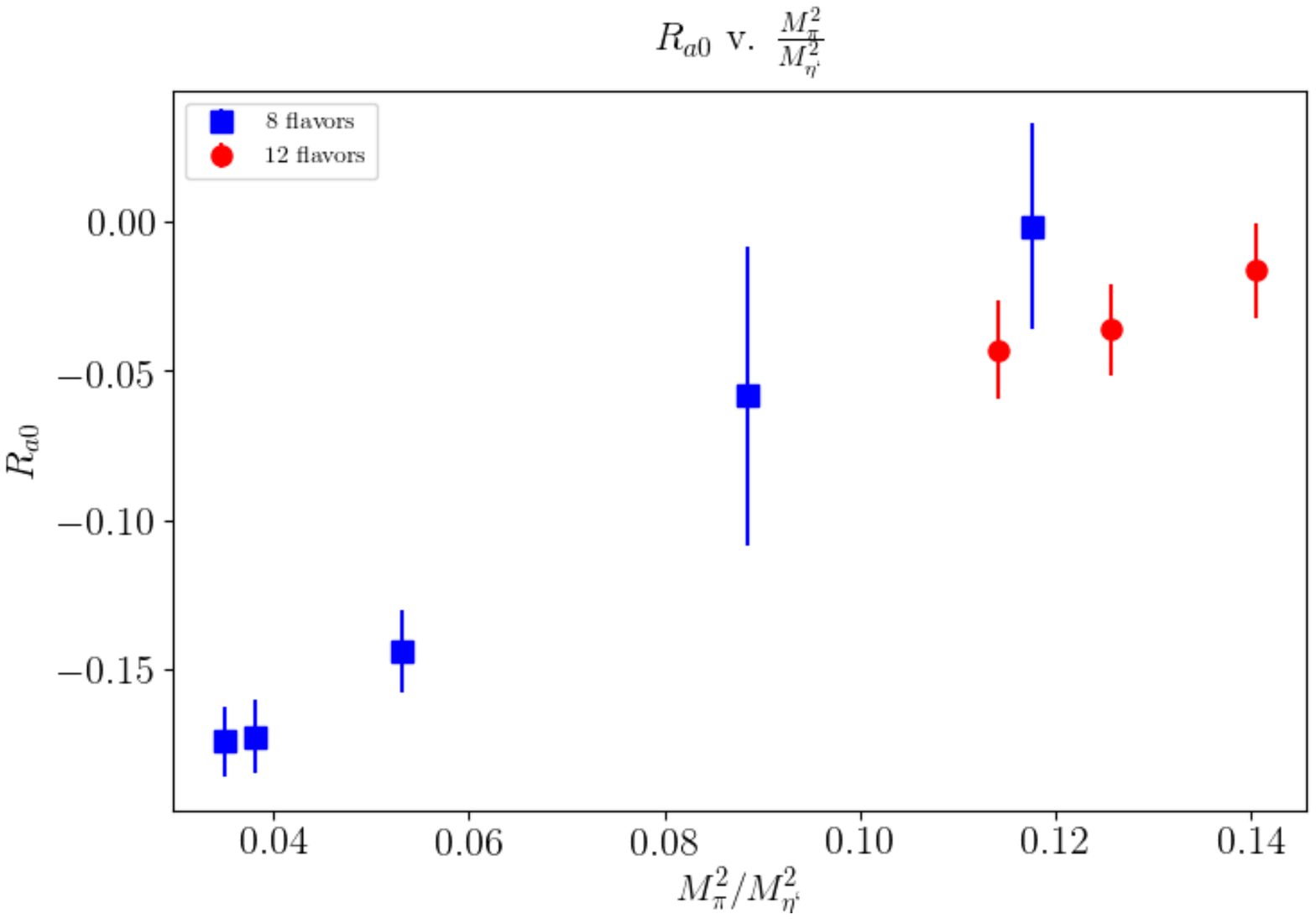}
\vskip-80pt
\caption{\label{fig:ratiosa}
$R_{a_0}$ for $N_f$=8 (squares) and 12 (circles)   versus 
$(M_\pi/M_{\etp})^2$.}
\end{figure}

Given the lack of reliable ways to calculate the mass dependence of the meson masses and coupling, we followed a 
phenomenological approach for the unperturbed spectrum \cite{meuriceprd96}. 
In the limit $m_1=m_2$, $v=v_1=v_2$, $Xv^{N_f-2}$, $\ls v^2$ and $\la v^2$  can 
be eliminated in terms of the zeroth-order masses. Introducing the notations
\begin{eqnarray}
\label{eq:main}
\Delta_{\sigma}&\equiv& M_{\sigma}^2-M_\pi^2\nonumber\\
\Delta_{a0}&\equiv& M_{a0}^2-M_\pi^2\\
\Delta_{\etp}&\equiv& M_{\etp}^2-M_\pi^2,\nonumber
\end{eqnarray}
these relations can be written as 
\begin{eqnarray}
\label{eq:main}
Xv^{N_f-2}&=&\Delta_{\etp}\nonumber\\
\ls v^2&=&\Delta_{\sigma}+(1-2/N_f)\Delta_{\etp}\\
\la v^2&=&\Delta_{a0}-(2/N_f)\Delta_{\etp}.\nonumber 
\end{eqnarray}
We also introduced the dimensionless ratios \cite{meuriceprd96}:
\begin{eqnarray}
R_{\sigma}&\equiv& \ls v^2/M_{\etp}^2,\\
R_{a_0}&\equiv& \la v^2/M_{\etp}^2.
\end{eqnarray}
Numerically, dividing by $M_{\etp}^2$ has a small effect because $0.87<aM_{\etp}<1.025$ and it removes the explicit dependance on the lattice spacing. 
This approach fixes the three unknown quantities with three numerical inputs. Interesting regularities 
are found for $R_{\sigma}$ and $R_{a_0}$. As an order of magnitude we found that for small $m_f$, $R_{\sigma}\simeq 1-2/N_f$ and $R_{a_0} \simeq -2/N_f$ with small variations with the fermion mass. Figs. \ref{fig:ratios} and \ref{fig:ratiosa}.
indicate that mass dependence has regularities that one should try to understand analytically. $R_\sigma$ decreases with $(M_\pi/M_{\etp})^2$ while $R_{a_0}$ increases. The order of magnitudes of the changes in $R_\sigma$ and $R_{a_0}$ are roughly the same as those of $(M_\pi/M_{\etp})^2$. 

This empirical data shows that  in order to fit the spectrum, the quartic couplings used in the tree-level mass formulas need to be tuned as  the symmetry breaking term changes. In a similar way, if we change $v_1$ and $v_2$ significantly, we also need to adjust the 
quartic coupling. For this reason, we will develop a perturbative approach where these adjustments would appear at second order.

\section{Perturbative splittings}
\label{sec:pert}

In this section we consider infinitesimal perturbation from the limit where $v_1=v_2=v$ with $v$ the vacuum expectation of $S_0$ for the one mass case. In the following, we work at first order in $v_2-v_1$  and the symbol $\simeq$ denotes equalities valid up to corrections of order $(v_2-v_2)^2$. Notice that the mixing terms $M_{S_{08}}^2$ and $M_{P_{08}}^2$ are of first order and have effects of second order on the mass eigenstates of the 2 by 2 matrix of singlets. Consequently mixing will be ignored in this section. 

\subsection{Adjoint splittings}
At first order in $v_2-v_1$, it is possible to express the splittings among the four masses of the scalars and pseudoscalars corresponding to  the decomposition of the $SU(N_f)_V$ adjoint given in Eq. (\ref{eq:split}), in terms of the unperturbed masses. 
For instance, 
\begin{eqnarray}
M_{\pi_{hh}}^2 - M_{\pi_{ll}}^2 &= &(v_2^2 - v_1^2)\Big(\frac{\lambda_{a0}}{2} +  \frac{X}{N_f}v_1^{N_1 - 2}v_2^{N_2 - 2}\Big)\nonumber \\
&\simeq & \frac{(v_2 - v_1)}{v}\Big( \lambda_{a0}v^2+  \frac{2X}{N_f}v^{N_f-2}\Big).
\end{eqnarray}

Using these relations we obtain 
\beq
M_{\pi_{hh}}^2 - M_{\pi_{ll}}^2\simeq  \frac{(v_2 - v_1)}{v}\Delta_{a0}.
\enq
Proceeding similarly, we obtain at first order: 
\begin{eqnarray}
M_{\pi_{lh}}^2 -M_{\pi_{ll}}^2 &\simeq &\frac{1}{2}(M_{\pi_{hh}}^2 - M_{\pi_{ll}}^2)\\
M_{P_{88}}^2 -M_{\pi_{ll}}^2 &\simeq &\frac{N_1}{N_f}(M_{\pi_{hh}}^2 - M_{\pi_{ll}}^2).\
\end{eqnarray}

It is possible to treat the scalar splittings in a completely similar way. The results are 
\begin{eqnarray}
M_{a0_{hh}}^2 - M_{a0_{ll}}^2&\simeq&  \frac{(v_2 - v_1)}{v}(3\Delta_{a0}-\frac{8}{N_f}\Delta_{\etp})\nonumber\\
M_{a0_{lh}}^2 -M_{a0_{ll}}^2 &\simeq &\frac{1}{2}(M_{a0_{hh}}^2 - M_{a0_{ll}}^2)\\
M_{S_{88}}^2 -M_{a0_{ll}}^2 &\simeq &\frac{N_1}{N_f}(M_{a0_{hh}}^2 - M_{a0_{ll}}^2).\nonumber
\end{eqnarray}

In order to have the expected splittings for $M_\pi^2$ where the pions containing lighter hyperfermions are lighter, we need to have $v_2>v_1$ because $\Delta_{a_0}>0$ (see Table \ref{tab:del}).  However, this choice implies that the various $\bao$ appear in the inverse order 
($M_{a0_{hh}}^2<M_{a0_{hl}}^2<M_{a0_{ll}}^2$) 
because $\frac{3\Delta_{a_0} - (8/N_f)\Delta_{\eta'}}{M_{\eta'}^2}<0$ (see Table \ref{tab:del}) for all the LatKMI datasets \cite{Aoki:2016wnc,Aoki:2017fnr}. The negative sign is clearly related to the lightness of the $a_0$ compared to QCD. 

\begin{table}[h]
\begin{tabular}{|c|c|c|c|}
\hline
\hline
$N_f$ & $(a m_f)$ &  $\frac{\Delta_{a_0}}{M_{\eta'}^2}$ & $\frac{3\Delta_{a_0} - (8/N_f)\Delta_{\eta'}}{M_{\eta'}^2}$\\\hline
8 & 0.012 & 0.0584(74) &  -0.669(71) \\\hline 
8 & 0.015 & 0.0644(78) &  -0.724(81) \\\hline 
8 & 0.02 & 0.0885(95) &  -0.640(70) \\\hline 
8 & 0.03 &  0.160(41) & -0.38(16) \\\hline 
8 & 0.04 &  0.214(28) &  -0.22(11) \\\hline 
12 & 0.04 &  0.0854(94) & -0.225(52) \\\hline 
12 & 0.05 & 0.1079(65) & -0.163(51) \\\hline 
12 & 0.06 & 0.1124(73)& -0.261(65) \\\hline 
\end{tabular}
\caption{Values  $\frac{\Delta_{a_0}}{M_{\eta'}^2}$ and $\frac{3\Delta_{a_0} - (8/N_f)\Delta_{\eta'}}{M_{\eta'}^2}$ using the LatKMI data \cite{Aoki:2016wnc,Aoki:2017fnr}.}
\label{tab:del}
\end{table}

The reasons for the inversion are clearly visible in Eq. (\ref{eq:scpsc}). Assuming $v_2>v_1$ to get the standard ordering for the pseudoscalars, we see that 
$\lambda_{a_0} <0$ makes $\lambda_{a_0}v_2^2$ more negative for 
$M_{a0_{hh}}^2$. In addition, the anomaly term for $M_{a0_{hh}}^2$ 
has larger powers of $v_1$ and lower powers of $v_2$ than $M_{a0_{ll}}^2$ 
and the coupling is positive so again it inverts the ordering. 
For the pseudoscalars, the coupling is negative so it is in normal order.

\subsection{Parametrization of $v_1$ and $v_2$}
So far we have not discussed how $v_1$ and $v_2$ differ from $v$ and from each other. 
In general, it is possible to select a path leaving the $v_1=v_2$ line in the $(v_1,v_2)$ plane 
in such a way that some condition is satisfied. An example of  condition  that we may like to impose is $\delta M_{\pi_{ll}}^2=0$. 
In the linear approximation this implies a linear relation between the changes. 

At first order, we will  use the parametrization: 
\beq
v_i=v(1+\epsilon g_i), 
\enq
for $i=1,\ 2$ and $g_i$ to be determined. 
At first order in $\epsilon$, we have the mass variations with respect to the unperturbed case
\beq
\delta M^2_{mes.}=\epsilon\sum_{i=1}^2g_iv\left. \frac{\partial M^2_{mes.}}{\partial v_i}\right\rvert _{v_1=v_2=v} .
\enq
where $mes.$ is any of the 10 mesons states. This subscript will be dropped in the following. 
The expressions
\beq
d_i\equiv v\left. \frac{\partial M^2}{\partial v_i}\right\rvert _{v_1=v_2=v},
\enq
are evaluated at zeroth-order and are functions of  $Xv^{N_f-2}$, $\ls v^2$ and $\la v^2$. This implies that the gradients can be expressed as 
\beq
d_i=d_{i\sigma}\Delta_{\sigma}+d_{ia0}\Delta_{a0}+d_{i\etp}\Delta_{\etp}.
\enq
The values of these coefficients are listed in Tables \ref{tab:der1} and \ref{tab:der2} for each of the meson states.
\vskip5pt
\begin{table}[h]
\begin{tabular}{|c|c|c|c|c|}
\hline
& $d_{1a0}$ & $d_{1\sigma}$ & $d_{1\eta'}$ \\
\hline
$\delta M_{\pi_{ll}}^2$ & $\frac{N_2}{N_f}$ & $\frac{N_1}{N_f}$ & 0  \\ \hline
$\delta M_{\pi_{lh}}^2$ & $\frac{N_f-2N_1}{N_f 2}$ & $\frac{N_1}{N_f}$ & 0 \\ \hline
$\delta M_{\pi_{hh}}^2$ & $- \frac{N_1}{N_f}$ & $\frac{N_1}{N_f}$ & 0  \\ \hline
$\delta M_{P88}^2$ & $\frac{N_2-N_1}{N_f}$ & $\frac{N_1}{N_f}$ & 0  \\ \hline
$\delta M_{P00}^2$ & $0$ & $\frac{N_1}{N_f}$ & $\frac{N_1(N_f-2)}{N_f}$ \\ \hline
$\delta M_{a0_{ll}}^2$ & $3-\frac{N_1}{N_f}$ & $\frac{N_1}{N_f}$ & $\frac{2(N_1-4)}{N_f}$\\\hline
$\delta M_{a0_{lh}}^2$ & $\frac{3}{2}-\frac{N_1}{N_f}$ & $\frac{N_1}{N_f}$ & $\frac{2(N_1-2)}{N_f}$\\\hline
$\delta M_{a0_{hh}}^2$ & $-\frac{N_1}{N_f}$ & $\frac{N_1}{N_f}$ & $2\frac{N_1}{N_f}$\\\hline
$\delta M_{S88}^2$ & $\frac{4N_2}{N_f}-1$ & $\frac{N_1}{N_f}$ & $\frac{2(N_1N_f-4N_2)}{N_f^2}$\\\hline
$\delta M_{S00}^2$ & 0 & $\frac{3 N_1}{N_f}$ & $-\frac{N_1(N_f^2 - 6N_f + 8)}{N_f^2}$\\\hline
\end{tabular}
\caption{\label{tab:der1}Coefficients $d_{1a0}$,  $d_{1\sigma}$ and  $d_{1\eta'}$}
\end{table}
\begin{table}[h]
\begin{tabular}{|c|c|c|c|c|}
\hline
& $d_{2a0}$ & $d_{2\sigma}$ & $d_{2\eta'}$ \\
\hline
$\delta M_{\pi_{ll}}^2$ & $-\frac{N_2}{N_f}$ & $\frac{N_2}{N_f}$ & 0  \\ \hline
$\delta M_{\pi_{lh}}^2$ & $\frac{N_f-2N_2}{N_f 2}$ & $\frac{N_2}{N_f}$ & 0 \\ \hline
$\delta M_{\pi_{hh}}^2$ & $\frac{N_1}{N_f}$ & $\frac{N_2}{N_f}$ & 0  \\ \hline
$\delta M_{P88}^2$ & $\frac{N_1-N_2}{N_f}$ & $\frac{N_2}{N_f}$ & 0  \\ \hline
$\delta M_{P00}^2$ & $0$ & $\frac{N_2}{N_f}$ & $\frac{N_2(N_f-2)}{N_f}$ \\ \hline
$\delta M_{a0_{ll}}^2$ & $-\frac{N_2}{N_f}$ & $\frac{N_2}{N_f}$ & $\frac{2N_2}{N_f}$\\\hline
$\delta M_{a0_{lh}}^2$ & $\frac{3}{2}-\frac{N_2}{N_f}$ & $\frac{N_2}{N_f}$ & $\frac{2(N_2-2)}{N_f}$\\\hline
$\delta M_{a0_{hh}}^2$ & $3-\frac{N_2}{N_f}$ & $\frac{N_2}{N_f}$ & $\frac{2(N_2-4)}{N_f}$\\\hline
$\delta M_{S88}^2$ & $\frac{3N_1-N_2}{N_f}$ & $\frac{N_2}{N_f}$ & $\frac{2(N_2N_f-4N_1)}{N_f^2}$\\\hline
$\delta M_{S00}^2$ & 0 & $\frac{3N_2}{N_f}$ & $-\frac{N_2(N_f^2 - 6 N_f + 8)}{N_f^2}$\\\hline
\end{tabular}
\caption{\label{tab:der2}Coefficients $d_{2a0}$,  $d_{2\sigma}$ and  $d_{2\eta'}$.}
\end{table}

\begin{figure}[h]
\includegraphics[width=8.5cm]{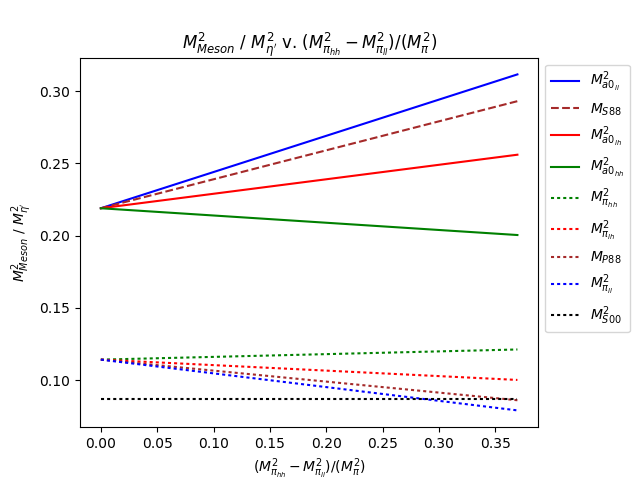}
\caption{\label{fig:split}
The spectrum of $M_{meson}^2$ in $M_{\etp}^2$ units, in the linear approximation for $N_1=2$ and $N_2=10$ versus $(M_{\pi_{hh}}^2 - M_{\pi_{ll}}^2)/M^2_\pi$. The unperturbed spectrum, which includes the values of 
$M_{\etp}^2$ and $M^2_\pi$ used in the graph come from the LatKMI data  discussed in the text for $N_f=12$ and $am_f=0.04$.}
\end{figure}

For practical applications it may be useful to choose the relation between $g_1$ and $g_2$ in such a way that some mass stays constant. 
The case $\delta M_\sigma^2=0$ is special because $d_1/d_2=N_1/N_2$ and consequently we can take $g_1/g_2=-N_2/N_1$.  This also implies
that  $\delta M_{\eta'}^2=0$. 
This choice is illustrated in Fig. \ref{fig:split}. 
We remind that at first order, mixings are neglected and so $M_\sigma \simeq M_{S00}$ and $M_{\eta'}\simeq M_{P00}$. 
Another interesting choice is $\delta M_{\pi_{ll}}^2=0$. However in this case the mass differences do not cancel in $d_1/d_2$ and the 
ratios of $g_1/g_2$ needs to be adjusted separately for different data sets. 
\section{Conclusions}
In summary, we have calculated the tree level spectrum for a linear model corresponding to a symmetry breaking with two different masses 
with adjustable multiplicities. We developed a perturbative expansion in the mass difference $(M_{\pi_{hh}}^2 - M_{\pi_{ll}}^2)$ which provides simple results for the 
ratios of differences of masses squared  (1/2 and $N_1/N_f$) and have similar structure for the adjoint of the scalar and pseudoscalars. However, 
when we impose the familiar ordering for the pseudoscalars ($M_{\pi_{ll}}^2<M_{\pi_{hh}}^2$) we obtained the inverse ordering for the 
scalars ($M_{a0_{ll}}^2>M_{a0_{hh}}^2$).

The two possible reasons for the inversion can be seen in Eq. (\ref{eq:scpsc}) and are $\lambda_{a_0}<0$ and the relative sign of the 
anomaly term in the spectrum. 
This inversion prediction could be verified or falsified by ongoing multiflavors simulations. 
The verification would be a surprising and interesting result. 
Numerical disagreement with the inversion would lead us to reconsider 
the two underlying reasons for the inversion.
It seems clear that if $m_2>>m_1$, the normal order should be restored. It is possible that 
the connection between the two regimes could be understood using the radiative corrections or sudden vacuum changes.

The inversion could be tested for instance with $N_1=2$ and $N_2=6$ with $a m_1=0.012$ and $am_2=0.015$ which are 
mass parameters used by LatKMI. The masses are small enough to have 
a clearly negative $\lambda_{a0}$ and the relative mass difference seems small enough to avoid large nonlinear corrections. 

In QCD, the $a_0$ can decay into $\eta\pi$ and is sometimes considered as a more complicated degree of freedom \cite{PhysRevD.77.094004}. However, in the single mass situation with $N_f=8$ or 12, where $\eta$ and $\pi$ are degenerate, the $a_0$ of the LatKMI data is light enough to forbid the on-shell process. We expect this property to remain valid with the small mass difference 
suggested above. This should make the lattice analysis simpler than in QCD. 

More generally our work should be considered as an encouragement to calculate spectra with 
two not so different masses and test model calculations of  the effects of the mass difference with a reliability comparable to what can be done with chiral perturbation 
theory for QCD. 

\acknowledgments
We thank A. Gasbarro and O. Witzel for discussions. 
This research was supported in part  by the Department of Energy 
under Award Numbers DOE grant DE-SC0010113. 
   \bibliography{centralmacbibb.bib}
\end{document}